\documentclass[aps,prd,twocolumn,superscriptaddress,preprintnumbers]{revtex4-1}%
\usepackage{epsf,epsfig}
\usepackage[utf8]{inputenc}
\usepackage{amssymb,amsmath,amsfonts}
\usepackage{color}
\usepackage{slashed}
\usepackage{braket}
\usepackage[colorlinks=true]{hyperref}  
\hypersetup{
    bookmarks=true,         
    unicode=false,          
    pdftoolbar=true,        
    pdfmenubar=true,        
    pdffitwindow=false,     
    pdfstartview={FitH},    
    colorlinks=true,       
    linkcolor=magenta, 
    citecolor=blue,        
    filecolor=magenta,      
    urlcolor=cyan           
} 

\newcommand{\bea}{\begin{eqnarray}}
\newcommand{\eea}{\end{eqnarray}}
\newcommand{\ba}{\begin{eqnarray}}
\newcommand{\ea}{\end{eqnarray}}
\newcommand{\rh}{r_{\rm h}}
\newcommand{\nn}{\nonumber \\}

\newcommand{\beq}{\begin{equation}}
\newcommand{\eeq}{\end{equation}}
\newcommand{\beqa}{\begin{eqnarray}}
\newcommand{\eeqa}{\end{eqnarray}}
\newcommand{\beqar}{\begin{eqnarray*}}
\newcommand{\eeqar}{\end{eqnarray*}}

\newcommand{\eg}{{\it e.g.,}\ }
\newcommand{\ie}{{\it i.e.,}\ }

\newcommand{\fin}{f_\infty}

\usepackage{color}

\newcommand{\req}[1]{(\ref{#1})} 

\begin{document}

\title{Geometric Inflation}
\author{Gustavo Arciniega}
\email{gustavo.arciniega@gmail.com}
\affiliation{Departamento de F\'isica de Part\'iculas, Instituto Galego de F\'isica das Altas Enerx\'ias (IGFAE), Universidade de Santiago de Compostela, E-15782 Santiago de Compostela, Spain}

\author{Pablo Bueno}
\email{pablo.bueno@cab.cnea.gov.ar}
\affiliation{Instituto Balseiro, Centro At\'omico Bariloche, S. C. de Bariloche, R\'io Negro, R8402AGP, Argentina}

\author{Pablo A. Cano}
\email{pablo.cano@uam.es}
\affiliation{Instituto de F\'isica Te\'orica UAM/CSIC,
	C/ Nicol\'as Cabrera, 13-15, C.U. Cantoblanco, 28049 Madrid, Spain}

\author{\\ Jos\'e D. Edelstein}
\email{jose.edelstein@usc.es}
\affiliation{Departamento de F\'isica de Part\'iculas, Instituto Galego de F\'isica das Altas Enerx\'ias (IGFAE), Universidade de Santiago de Compostela, E-15782 Santiago de Compostela, Spain}

\author{Robie A. Hennigar} 
\email{rhennigar@mun.ca}
\affiliation{Department of Mathematics and Statistics, Memorial University of Newfoundland, St. John’s, Newfoundland and Labrador, A1C 5S7, Canada}

\author{Luisa G. Jaime}
\email{luisa.jaime1@gmail.com}
\affiliation{Instituto de Ciencias Nucleares, Universidad Nacional Aut\'onoma de M\'exico, A.P. 70-543 CDMX 04510, Mexico}


\preprint{IFT-UAM/CSIC-18-133}

\begin{abstract}
We argue that the presence of an inflationary epoch is a natural, almost unavoidable, consequence of the existence of a sensible effective action involving an infinite tower of higher-curvature corrections to the Einstein-Hilbert action. No additional fields besides the metric are required. We show that a family of such corrections giving rise to a well-posed cosmological evolution exists and automatically replaces the radiation-dominated early-universe Big Bang by a singularity-free period of exponential growth of the scale factor, which is gracefully connected with standard late-time $\Lambda$CDM cosmology. The class of higher-curvature theories giving rise to sensible cosmological evolution share additional remarkable properties such as the existence of Schwarzschild-like non-hairy black holes, or the fact that, just like for Einstein gravity, the only degrees of freedom propagated on the vacuum are those of the standard graviton.
\end{abstract}

\maketitle

The Einstein-Hilbert action is expected to be the first in an infinite series of higher-curvature terms which become relevant at sufficiently high energy scales. This is, for instance, what String Theory predicts, corrections being weighted by powers of $\alpha'$ or the Planck length, \eg \cite{Gross:1986mw,Gross:1986iv,Grisaru:1986vi,Bergshoeff:1989de,Green:2003an,Frolov:2001xr}. Naively, one would expect the terms appearing in the corresponding stringy four-dimensional effective actions to give rise to sensible modifications of Einstein's gravity predictions, for example, yielding second-order equations for cosmological evolution. However, it is well-known that this is not the case in general ---for instance, one can get the standard $R^2$ Starobinsky model \cite{Starobinsky:1980te} corrected by a Weyl$^2$ term \cite{Prue}, which would spoil the nice behavior of the former. The idea is that such ``inconsistencies'' would be healed if one considers the full-fledged UV-complete theory; they would be artifacts resulting from truncating the series at some particular order in the parameter expansion. We do not control the exact mechanism by which, say, String Theory (or any other putative UV-complete description of gravity) deals with the infinite tower of effective higher-curvature terms at sufficiently high energies. Whatever it is, though, the net result must be to produce sensible dynamics.

It is not always the case that we have at our disposal a UV-complete theory from which we can derive the low energy dynamics in a top-down approach. Hence, an alternative attempt at capturing such higher energy effects would consist in considering effective actions which already satisfy the requirement of producing sensible dynamics by themselves at all orders. This is in line with recent approaches to scrutinize candidates for low energy effective actions by demanding the absence of physical inconsistencies such as unitarity or causality violation order by order \cite{Adams:2006sv,Camanho:2014apa}. We shall follow this bottom-up approach with a broad question in mind: if any meaningful description of the cosmological dynamics can be expressed as an effective action for the metric tensor, order by order, what would be its predictions?

In this letter we analyze this question by constructing an infinite family of higher-curvature corrections to the Einstein-Hilbert action, selected by the criterium that they give rise to a well-posed cosmological problem. Interestingly, we find that the standard radiation-dominated decelerating early-universe predicted by Einstein's gravity (in the absence of additional mechanisms) is generically replaced by an inflationary era of exponential growth of the scale factor. For these ``geometric inflation'' models, the problem would be explaining a universe without inflation rather than the opposite. The effect of the higher-curvature terms becomes increasingly irrelevant as $a(t)$ grows and, eventually, the evolution smoothly transits to standard post-inflationary $\Lambda$CDM cosmology.
We stress that, in this setup, inflation takes place regardless of any extra (scalar) fields. This seems also relevant in the context of the recent discussions about inflationary models and the swampland conjectures \cite{Obied:2018sgi,Agrawal:2018own}.

The class of theories satisfying the well-posed cosmology criterium ---whose first representative was constructed in \cite{Arciniega:2018fxj}--- is a subclass of a broader, recently identified, family \cite{PabloPablo,Hennigar:2016gkm,PabloPablo2,Hennigar:2017ego,PabloPablo3,Ahmed:2017jod,PabloPablo4}, characterized by additional remarkable properties such as possessing second-order equations of motion on maximally symmetric backgrounds, as well as the existence of non-hairy generalizations of the Schwarzschild black hole (see below). 

Let us now present the results and postpone all further comments to the closing discussion section.

\vskip2mm
\noindent
\textbf{The theory}
\vskip1mm

\noindent
At every order in curvature, and up to multiplicities, there exists a density $\mathcal{R}_{(n)}$ constructed from contractions of the metric and the Riemann tensor such that the theory defined by the action 
\begin{equation}\label{theo}
S=\int \frac{d^4x \sqrt{|g|}}{16\pi G}\left\{-2\Lambda+R+\sum_{n=3}^{\infty}\lambda_n L^{2n-2}\mathcal{R}_{(n)}\right\}\, ,
\end{equation}
\noindent 
satisfies the following properties: i) it possesses second-order linearized equations around any maximally symmetric background; ii) it admits bona-fide non-hairy generalizations of the Schwarzschild(-AdS) black hole (and, more generally, of the Einstein gravity Taub-NUT/bolt solutions) characterized by a single function, $g_{tt}g_{rr}=-1$, and whose thermodynamic properties can be accessed in a fully analytic fashion; iii) it possesses a well-posed cosmological initial-value problem, namely, it admits cosmological Friedmann-Lema\^itre-Robertson-Walker (FLRW) solutions
\begin{equation}\label{FLRW}
ds^2=-dt^2+a(t)^2 \left(\frac{dr^2}{1-k r^2}+r^2 d\Omega^2 \right)\, ,
\end{equation}
where the associated generalized Friedmann equations for the scale factor $a(t)$ are second-order. 

 Notice that in \req{theo} we have introduced dimensionless couplings $\lambda_n$, as well as a new energy scale $\sim L^{-1}$, which we will tacitly assume to be below the Planck scale, $L_{\rm Pl}^{-1}$. If $L^{-1} \ll L^{-1}_{\rm Pl}$ and $\lambda_3 \neq 0$, the theory is afflicted by causality issues, which can be seen as a fingerprint of the presence of an infinite tower of massive higher spin particles \cite{Camanho:2014apa}. We will assume that $L^{-1} \lesssim L^{-1}_{\rm Pl}$ and will be more explicit on the values of $\lambda_n$ below.

The discovery of four-dimensional theories satisfying requirements i) and ii) was ignited by the construction of Einsteinian cubic gravity \cite{PabloPablo,Hennigar:2016gkm,PabloPablo2}. By now, the existence of a family of theories satisfying such properties ---including terms with arbitrarily high orders in curvature--- and their connection to previously known (and some unknown) higher-dimensional theories such as Lovelock \cite{Lovelock1,Lovelock2} or Quasi-topological gravities \cite{Quasi2,Quasi,Dehghani:2011vu,Cisterna:2017umf} is well established \cite{Hennigar:2017ego,PabloPablo3,Ahmed:2017jod,PabloPablo4,Feng:2017tev,Hennigar:2017umz,Li:2017ncu,HoloECG,Bueno:2018uoy,Cisterna:2018tgx}. The observation that a subclass of such theories also satisfies iii) ---at least at cubic order--- was put forward in \cite{Arciniega:2018fxj}, and subsequently extended to the quartic and quintic orders in \cite{Cisterna:2018tgx}. The cubic representative turns out to involve a simple linear combination of the Einsteinian cubic gravity density $\mathcal{P}=12 R_{a\ b}^{\ c \ d}R_{c\ d}^{\ e \ f}R_{e\ f}^{\ a \ b}+R_{ab}^{cd}R_{cd}^{ef}R_{ef}^{ab}-12R_{abcd}R^{ac}R^{bd}+8R_{a}^{b}R_{b}^{c}R_{c}^{a}$ and a previously characterized \cite{Hennigar:2017ego} invariant $\mathcal{C}=R_{abcd}R^{abc}\,_{e}R^{de}-\frac{1}{4}R_{abcd}R^{abcd}R-2R_{abcd}R^{ac}R^{bd}+\frac{1}{2}R_{ab}R^{ab}R$, which is trivial when evaluated on a static and spherically symmetric ansatz; the exact combination reads $\mathcal{R}_{(3)}\propto\mathcal{P}-8\mathcal{C}$. In appendix \ref{densis} we present the explicit form of $\mathcal{R}_{(n)}$ up to $n=8$ as well as a rationale to construct these terms in general. 


\vskip2mm
\noindent
\textbf{Generalized Friedmann equations}
\vskip1mm

\noindent
When evaluated on a FLRW ansatz of the form \req{FLRW}, the full non-linear equations of \req{theo} reduce to a couple of second-order differential equations for the scale factor. Focusing on flat spatial metrics, $k=0$, the generalized Friedmann equations read \footnote{
In the general case, $k\neq 0$, the equations read \begin{align}
\Lambda-3F\left(h \right)&=-8\pi G \rho\, ,\\
\Lambda- \left[\frac{F'(h)}{H} \frac{dh}{dt} + 3 F(h) \right]&=+8 \pi G P \, ,
\end{align}
where we introduced $h\equiv \sqrt{H^2+k/ a^2}$}
\begin{align}\label{Fried}
3F(H)&=8\pi G \rho+\Lambda\, ,\\ \label{Fried2}
-\frac{\dot{H}}{H}F'(H)&=8 \pi G (\rho+P) \, ,
\end{align}
where
\begin{equation}\label{F}
F(H)\equiv H^2+L^{-2}\sum_{n=3}^{\infty} (-1)^n\lambda_n \left(LH\right)^{2n}\, ,
\end{equation}
and $F'(H)\equiv dF(H)/dH$. Here, $H\equiv \dot{a}/a$ is the usual Hubble parameter, and $\rho$ and $P$ are the density and pressure of a perfect fluid, whose energy-momentum tensor appears in the right-hand-side of the modified Einstein's equations.  Observe that if we set all the higher-curvature couplings to zero, $F(H)=(\dot a/a)^2$, $\dot{H} F'(H)/H=2 (\ddot a a -\dot a^2)/a^2$, and \req{Fried} and \req{Fried2} reduce to the usual Einstein-gravity versions. Note also that using both equations one obtains the conservation equation
$$
\frac{d\rho}{dt} + 3H(P+\rho) = 0 \,.
$$
Assuming that the fluid is composed of matter ($P_m=0$) and radiation ($P_r=\rho_r/3$), one gets the familiar expression for the energy density
\begin{equation}\label{densi}
\rho=\rho_m+\rho_r = \frac{3F(H_0)}{8\pi G} \left[\frac{\Omega_m}{a^3}+\frac{\Omega_r}{ a^{4}} \right]\, ,
\end{equation}
where $\Omega_m = \rho_m/\rho_c$ and $\Omega_r=\rho_r/\rho_c$. As in \cite{Arciniega:2018fxj}, we redefine the critical density as
$
\rho_c =3F(H)/(8\pi G)\, ,
$
in order to avoid possible misinterpretations of our models as some kind of dark fluid.

Interestingly enough, the cosmological evolution of this class of theories is entirely dictated by the function $F(H)$ \footnote{This is reminiscent of what happens for black holes in Lovelock gravity \cite{Camanho:2011rj}; indeed, Lovelock cosmologies display a similar structure \cite{Camanho:2015ysa}}, which is in turn controlled by $L^{-1}$ and the $\lambda_n$.
Ultimately, one would like to be able to compute those couplings from a putative UV-complete theory. In that respect, notice that there are ambiguities in the definition of the $\mathcal{R}_{(n)}$ appearing in (\ref{theo}): there exist distinct densities ${\cal R}_{(n)}^A$ and ${\cal R}_{(n)}^B$ differing by a third density ${\cal T}_{(n)}^{AB}$ with the property that ${\cal T}_{(n)}^{AB}$ makes no contribution to the field equations for the classes of metrics considered here (see appendix \ref{densis}). If we call $\lambda_n^A$ and $\lambda_n^B$ the gravitational couplings multiplying these two densities, the only effect in the cosmology will be the appearance of either coupling in (\ref{F}). That is, it is tantamount an ambiguity in the numerical value of the $n$-th gravitational coupling. We expect that this is fixed by either an explicit UV-completed theory or by further consistency-checks that the theory must overcome. In the meanwhile, we will extract generic conclusions from the very fact that there exists such a function and we will focus on two particular models for the sake of definiteness.

\vskip2mm
\noindent
\textbf{Explicit models}
\vskip1mm

\noindent
At this level, the gravitational couplings $\lambda_n$ are free parameters in our Lagrangian. This gives rise to a vast range of possible models, each one with their own particularities. There are, nonetheless, certain reasonable constraints that can be imposed. On the one hand, requiring positive-mass black holes to exist, would fix the sign of the first non-vanishing coupling \cite{PabloPablo4}, \eg $\lambda_3>0$. It is also convenient to choose $F(H)$ to be a bijective function. This avoids pathological situations such as the absence of solutions for high enough energy densities, or the appearance of singularities at points for which $F(H)$ reaches an extremum. The simplest way of satisfying these requirements consists in setting all odd-order couplings to zero and imposing the even ones to be positive, \ie $\lambda_{2k+1}=0$, $\lambda_{2k}>0$, for all $k\in \mathbb{Z}^+$. The odd ones can be safely included if they are negative (except for $\lambda_3$), or positive but sufficiently small. Besides this, we can also choose the relative values of the different couplings. For concreteness, in this letter we will consider two models to be compared with the standard $\Lambda$CDM one, namely
\begin{eqnarray}
\label{m1}
&\text{Model 1:} &\quad \lambda_{2k+1}=0\, ,\quad \lambda_{4+2k}=\lambda_4/ k!\, , \quad k\in \mathbb{Z}^+\, ,\\ [0.4em]
\label{m2}
&\text{Model 2:} &\quad \lambda_3>0\, ,\quad \lambda_{n\geq 4}=(-1)^n \lambda_3 /(n-4)! \, .
\end{eqnarray} 
The above choices simplify the functional form of $F(H)$, making it possible to sum the corresponding infinite series (many other summable choices are possible). One finds $F(H)=H^2+\lambda_4 H^8L^6 e^{(HL)^4}$ for Model 1 and  $F(H)=H^2-\lambda_3L^4 H^6[1- (HL)^2 e^{(HL)^2}]$ for Model 2. With regards to $L^{-1}$, the most reasonable choice seems to be that it corresponds to some new scale below the Planck mass, but high enough to make the higher-curvature effects become negligible at late times.

\begin{figure}[ht] \hspace{-0.1cm}
\includegraphics[scale=0.75]{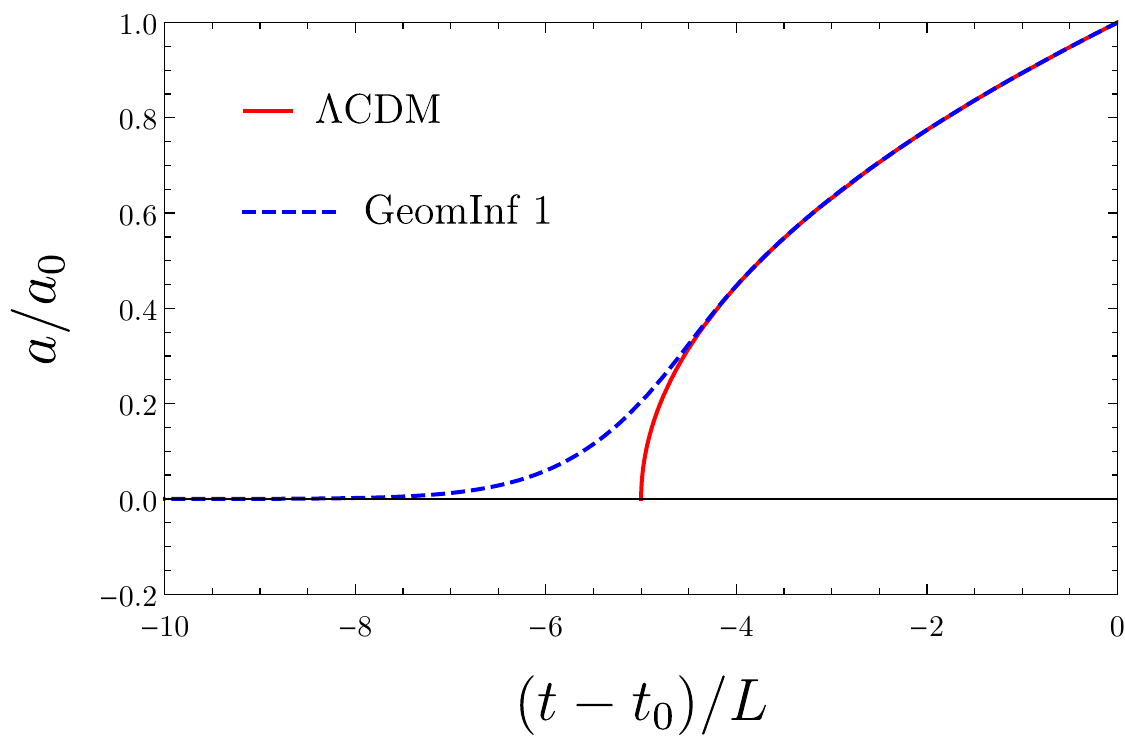}
\caption{ 
Scale factor $a(t)$ as a function of time as predicted by the standard $\Lambda$CDM model and  our geometric inflation Model 1 defined in \req{m1} (for clarity reasons, we omit the curve corresponding to Model 2 in this plot). The normalization of $a$ is arbitrary. Here, we have normalized it by $a_0$ corresponding to its value at some later time $t_0$ at which the higher-curvature corrections become negligible. In particular, we have fixed this value so that $H(t_0)=0.1 L^{-1}$. For the $\Lambda$CDM model, driven by Einstein-gravity dynamics, the Big Bang singularity is reached at $t-t_0=-5 L$. When the higher-curvature corrections are taken into account, the value $a=0$ is never reached at finite times, but only approached asymptotically as $t\rightarrow -\infty$.}
\label{fig.1}
\end{figure}

\begin{figure}[ht] \hspace{-0.1cm}
\includegraphics[scale=0.75]{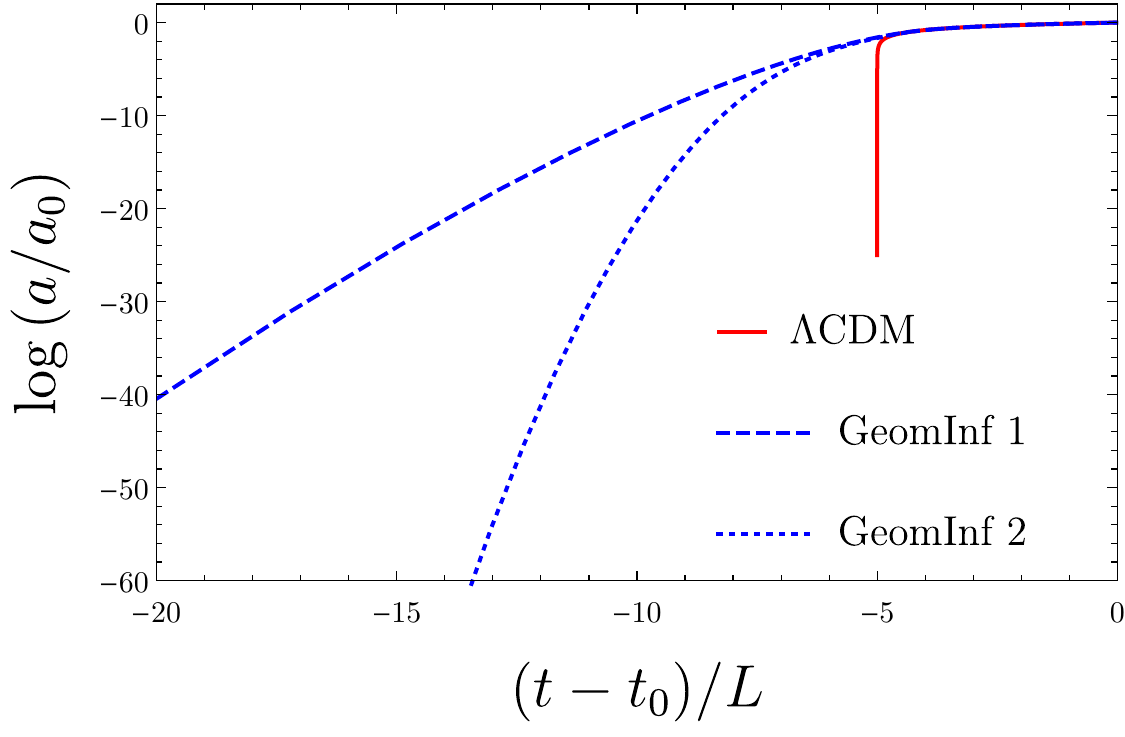}
\caption{ 
We plot $\log(a(t)/a_0)$.  In geometric inflation models, the Big Bang is replaced by a period of exponential growth, which is made evident.}
\label{fig.2}
\end{figure}

\begin{figure}[ht] \hspace{-0.2cm}
\includegraphics[scale=0.69]{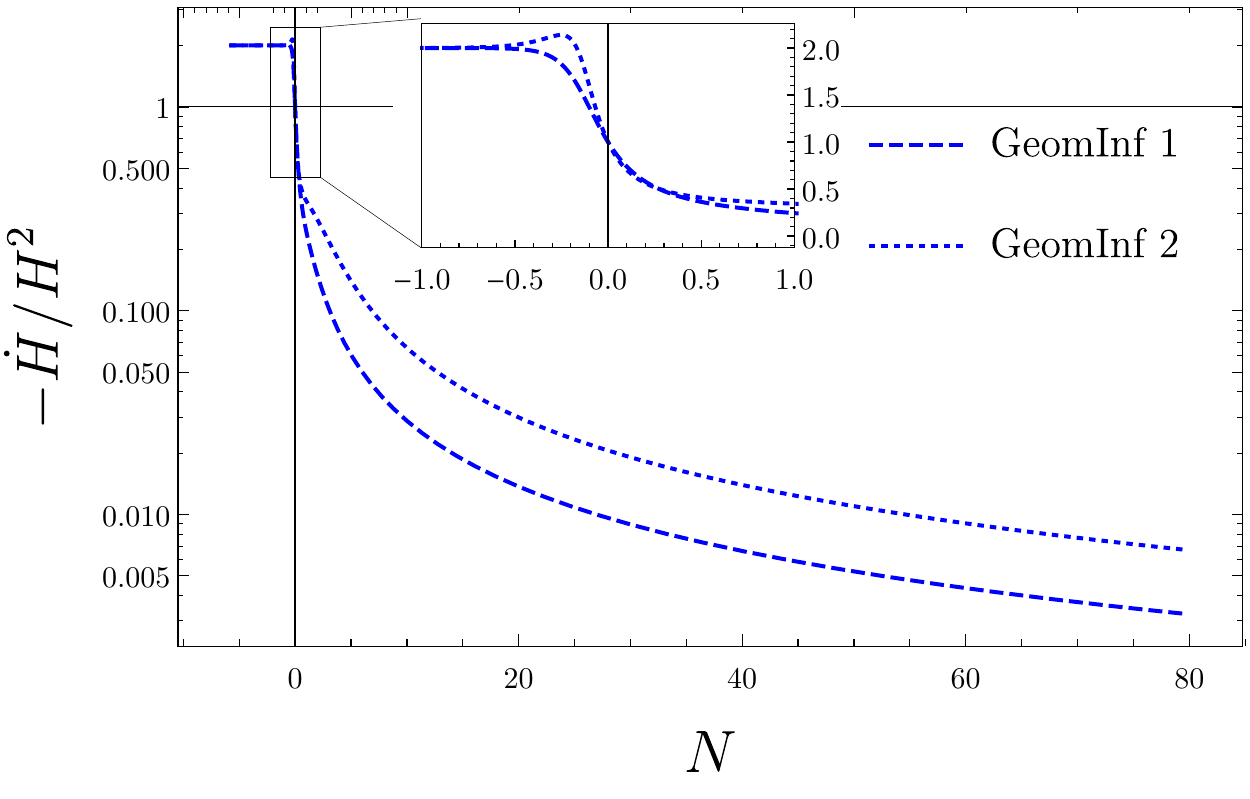}
\caption{ 
Slow-roll parameter $\epsilon=-\dot{H}/H^2$ as a function of the number of \emph{e-folds} $N=\log(a/a_f)$. The reference value $a_f$ is taken such that $\epsilon(a_f)=1$, meaning that inflation ends. (Inset) Zoom of the previous plot where we observe that there is a smooth transition between the Einstein gravity value for radiation-dominated universes, $\epsilon=2$, and the inflationary period, in which $\epsilon\rightarrow 0$. Geometric inflation provides a graceful exit to the inflationary period. }
\label{fig.3}
\end{figure}

In Figs.~\ref{fig.1}, \ref{fig.2} and \ref{fig.3} we denote these models ``GeomInf 1'' and ``GeomInf 2'', respectively.

\vskip2mm
\noindent
\textbf{Early universe cosmology}
\vskip1mm

\noindent
For small values of the scale factor, radiation dominates over matter and dark energy. In that regime, the standard Einstein gravity scaling is $a(t)\sim t^{1/2}$, which follows straightforwardly from \req{Fried} and \req{densi} for $F(H)=H^2$. Without additional mechanisms, this is obviously unsatisfactory from an inflationary perspective, as it predicts a negative acceleration $\ddot{a}/a\sim -1/(4t^2)$. The introduction of higher-order terms changes this behavior dramatically. Consider first the case in which we truncate the series in \req{theo} at a certain order in curvature $n=n_{\rm max}$, as in  \cite{Arciniega:2018fxj}. Neglecting all contributions but radiation, the generalized Friedmann equations can be solved parametrically as \footnote{In fact, a parametric solution can be written in general. The result is
\begin{equation}
a(p)=J(p)\, , \quad
t(p)=\int_{p_0}^p\frac{dq J'(q)}{ q J(q)}\, ,
\end{equation}
where $J$ is the function $J=P^{-1}\circ F$ and $p_0$ is such that $J(p_0)=1$.}
\begin{equation}\label{panam}
a(p)=a(p_0)\left(\frac{F(p_0)}{F(p)}\right)^{1/4}\!\!\!\!\! , \quad
t(p)=-\int_{p_0}^p\frac{dq}{4 q} \frac{F'(q)}{F(q)}\, .
\end{equation}
For a given $F(H)$, it is straightforward to obtain $a(t)$ from these equations. The limit $a\rightarrow 0$ implies $F(p)\rightarrow \infty$, which for the kind of models we are considering corresponds to $p\rightarrow \infty$. We find $F(p)\sim L^{2(n_{\rm max}-1)} p^{2n_{\rm max}}$ and $t(p)\sim n_{\rm max}/(2p)$, where the proportionality constants depend on the particular model. Putting this together in \req{panam}, we find
\begin{equation}
a(t)\sim t^{n_{\rm max}/2}\,  \quad \text{when} \quad t\rightarrow 0\, .
\end{equation}
Of course, this reduces to the Einstein gravity result for $n_{\rm max}=1$. Crucially, the introduction of higher-curvature terms changes the sign of the scale factor acceleration making it positive, namely $\ddot{a}/a\sim n_{\rm max}(n_{\rm max}-2)/(4t^2)$. Hence, positively accelerated expansion occurs provided
\begin{equation}\label{posiac}
\ddot{a}(t) > 0 ~\Leftrightarrow~ n_{\rm max}>2\, .
\end{equation}
The larger $n_{\rm max}$, the faster the acceleration at early times. This fact can also be expressed in terms of the slow-roll parameter, $\epsilon\equiv -\dot H/H^2=1-\ddot a/(a H^2)$, for which we find $\epsilon\sim 2/ n_{\rm max}$. Positive acceleration occurs for $0<\epsilon<1$, which of course translates into \req{posiac}.

On the other hand, when we only include a finite number of terms, the expansion is polynomial rather than exponential, and at some finite moment in the past a singularity would be reached when $a=0$, albeit the series truncation certainly ceases to be trustable when $a \sim L$. This suggests that, when dealing with early time cosmology, it is not consistent to truncate the higher-order terms, as they will become relevant before reaching the singularity. 
One can immediately see that when the full tower of higher-curvature terms is included, the scale factor will grow faster than any polynomial near $a=0$. Therefore, the expansion will be at least exponential. This can be verified explicitly for our two models in \req{m1} and \req{m2}. In particular, when $t\rightarrow-\infty$ we find $a(t)\sim e^{-(-3t/L)^{4/3}/4}$ for Model 1, and $a(t)\sim e^{-t^2/L^2}$  for Model 2. More generally, if $F(H)\sim e^{(HL)^w}$, one finds (up to multiplicative powers which are irrelevant for the order of magnitude)
\begin{equation}
a(t)\sim \exp\left[-\frac{1}{4}\left(\frac{4(1-w)t}{wL} \right)^{\frac{w}{w-1}} \right]\quad \text{if} ~ w>1\, ,
\end{equation}
and $a(t)\sim \exp \left[-e^{-4t/L}/4 \right]$ if $w=1$.
Observe that for these models, there is no notion of {\it origin of time} at which we can set $t=0$. Rather, $a\rightarrow 0$ is only reached asymptotically as $t\rightarrow -\infty$.

\vskip2mm
\noindent
\textbf{Geometric Inflation}
\vskip1mm

\noindent
Let us now analyze our two models in more detail. In order to do so, we solve the generalized Friedmann equations in each case. In Fig.~\ref{fig.1},
we plot the scale factor as a function of time for the standard $\Lambda$CDM model and Model 1 as defined in \req{m1}. The general intuitions observed in the previous section become manifest. In particular, we observe that the Big Bang singularity followed by a decelerating expansion predicted by Einstein's gravity for a radiation-dominated universe is replaced by an exponential growth of the scale factor which eventually connects with the Einstein gravity prediction at sufficiently late times, where the higher-curvature terms become increasingly irrelevant.

This behavior is even more manifest in the logarithmic plot of Fig.~\ref{fig.2}, where we also include Model 2 in the comparison.
Naturally, the exact details depend on the exact choice of parameters $\lambda_n$ and energy scale $L^{-1}$, but the general message is that an inflationary epoch appears to be  unavoidable, regardless of the specific model. It is also worth emphasizing that late-time cosmology can be essentially identical to the one predicted by Einstein gravity, as long as the new energy scale is high enough. Again, the exact point at which the crossover from the geometric inflation era to the standard $\Lambda$CDM-like phase occurs, depends on the exact details of each model, but a \emph{graceful exit} is automatically implemented in all models. This smooth transition is manifest in Figs.~\ref{fig.1} and \ref{fig.2}.

In Fig.~\ref{fig.3}, we plot the slow-roll parameter $\epsilon$ as a function of the number of \emph{e-folds}.
This is defined as $N=\log(a/a_f)$, where the reference value $a_f$ denotes the value at which inflation ends, defined so that $\epsilon(a_f)=1$.  For our class of theories, these two quantities can be written in terms of the function $F(H)$ defined in \req{F} as
\begin{equation}
N=\frac{1}{4}\log \left(\frac{F(H)}{F(H_f)} \right) \, , \quad \epsilon=\frac{4F(H)}{H F'(H)}\, .
\end{equation} 
When $\epsilon=0$, $H$ remains constant and the expansion is exponential ---corresponding to a pure de Sitter space. When $|\epsilon|\ll 1$, which is the case for most values of $N$ shown in the plot for our two models ---and continues to be the case as $N\rightarrow \infty$--- the growth is quasi-exponential. This is the typical behavior of standard slow-roll inflationary models (see, for example, \cite{Baumann:2009ds}). After the end of inflation is reached, the slow-roll parameter smoothly connects with the radiation-dominated Einstein gravity result, corresponding to $\epsilon=2$ ---see inset in Fig.~\ref{fig.3}.

As the $\epsilon(N)$ curves make manifest, our models implement all nice features of slow-roll models  without introducing any additional fields and/or potentials. The behavior of $\epsilon$ for large $N$ is $\epsilon\sim 1/(4N)$ for Model 1, and $\epsilon\sim 1/(2N)$ for Model 2. More generally, if $F(H)\sim e^{(HL)^w}$, we get $\epsilon\sim 1/(w N)$. However, observational quantities such as the spectral index will have a different dependence on $\epsilon$ with respect to the slow-roll models driven by scalars. For that, one would need to perform a thorough analysis of perturbations.

\vskip2mm
\noindent
\textbf{Discussion}
\vskip1mm

\noindent
Under the present proposal, the evolution of the Universe from inflation to late time acceleration can be explained by the bottom-up construction of an effective geometric field theory of gravity. We solely demand consistence of the theory order-by-order in the series of higher-curvature terms. We showed that if the full series is taken into consideration, and only in that case, then the expansion of the Universe at early times has no other choice than becoming exponential. As proved in \cite{Arciniega:2018fxj} for the cubic case, the evolution of the scale factor smoothly transits to standard post-inflationary $\Lambda$CDM cosmology.

The surprising performance of our theory in describing the accelerated expansion periods of the Universe is nothing but a first step towards building up a new cosmological framework. Further checks will represent an increasingly demanding challenge; for instance, the study of cosmological perturbations. Remarkably, a preliminary analysis shows that those also satisfy second-order dynamics \footnote{On general grounds, it is expected that higher-order time derivatives still appear at non-linear level in perturbation theory. One should study in that case whether there is a strong coupling problem, or if, on the contrary, the non-linear terms can be safely neglected.} ---this was anticipated in \cite{Cisterna:2018tgx} for the cubic theory--- which opens up the window for a more thorough study of observational tests such as the signatures in the Cosmic Microwave Background (CMB) \cite{Ade:2015xua}.

The features above described are achieved without the introduction of any additional fields besides the metric. As opposed to other alternatives studied in the cosmological context ---see \eg \cite{Sotiriou:2008rp,Horndeski:1974wa}---, our theories cannot be thought of as  scalar-tensor models. In fact, on maximally symmetric backgrounds they only propagate the usual transverse and traceless graviton of General Relativity so, in a sense, they are as ``pure-metric'' as it gets amongst higher-curvature theories (at least in four dimensions). On this respect, our proposal sheds a different light on discussions such as the role of reheating ---it is not clear at the moment how to deal with it--- and the status of the string swampland conjecture \cite{Obied:2018sgi,Agrawal:2018own} under the present geometrical scheme ---perhaps scalar fields play a less important role in the early universe than our current expectations suggest.

A better understanding on the multiplicities of the $\mathcal{R}_{(n)}$ densities and the related ambiguities in the actual numerical values of the gravitational couplings $\lambda_n$ seems necessary. It was shown in \cite{Cisterna:2018tgx} that the cubic density $\mathcal{R}_{(3)}$ constructed in \citep{Arciniega:2018fxj} is the dimensional reduction of so-called five dimensional quasi-topological gravity \cite{Quasi2,Quasi}. On the one hand, it is intriguing to explore whether there is a network connecting these classes of theories in different spacetime dimensions that may ultimately provide hints of a hidden mathematical structure. A complementary viewpoint would be to further scrutinize the $\mathcal{R}_{(n)}$ densities to better understand whether the triviality of the ambiguities ---when evaluated on spherically symmetric and cosmological ans\"atze--- is due to some sort of integrable structure. The value of the effective couplings, $\lambda_n$, can be otherwise constrained by using observations (SNIa \cite{Riess:1998cb}, \cite{Perlmutter:1998np}, $H_0$ \cite{Farooq:2013hq}  , BAO's \cite{Delubac:2014aqe}, \cite{Alam:2016hwk}).

The very fact that a purely geometric mechanism ---resting on generic aspects of a low energy effective approach to the gravitational interaction--- triggers inflation is tantalizing. More investigations are necessary in order to determine whether this idea entails a research program that resembles a long avenue or a dead-end street. We expect to contribute further to put together more pieces of this exciting puzzle.\\

\vskip2mm
\noindent
\textbf{Acknowledgements}
\vskip1mm

\noindent
We would like to thank Javier M. Mag\'an and Jos\'e Beltr\'an for useful discussions. 
The work of PB was supported by the Simons foundation through the It From Qubit Simons collaboration.
The work of PAC is funded by Fundaci\'on la Caixa through a ``la Caixa - Severo Ochoa'' International pre-doctoral grant  and partially by the MINECO/FEDER, UE grant FPA2015-66793-P, and the ``Centro de Excelencia Severo Ochoa'' Program grant SEV-2016-0597.
The work of JDE is supported by MINECO FPA2014-52218 and FPA2017-84436-P, Xunta de Galicia ED431C 2017/07, FEDER, and the Mar\'\i a de Maeztu Unit of Excellence MDM-2016-0692.
The work of RAH is supported by the Natural Sciences and Engineering Research Council of Canada through the Banting Postdoctoral Fellowship programme.

\onecolumngrid  \vspace{1cm} 
\begin{center}  
{\Large\bf Appendices} 
\end{center} 
\appendix 

\section{Form of the densities up to $n=8$}
\label{densis}


Here we present explicit examples of higher-order Lagrangians that lead to the equations of motion presented in the main text. The full gravitational part of the action reads
\begin{equation}
S=\int \frac{d^4x \sqrt{|g|}}{16\pi G}\left\{-2\Lambda+R+\sum_{n=3}^{\infty}\lambda_n L^{2n-2}\mathcal{R}_{(n)}\right\}\, ,
\end{equation}
where the ${\cal R}_{(n)}$ are chosen to meet the following criteria:
\begin{enumerate}
\item They are normalized so that the embedding equation of the theory reads $ -L^2\Lambda/3 - \fin + \sum_{n=3} \lambda_n \fin^n = 0$ for maximally symmetric vacua with curvature $-\fin/L^2$. 
\item The theory admits non-hairy, spherically symmetric black hole (and more generally, Taub-NUT/Bolt) solutions as described in appendix~\ref{sec:schwHoles}.
\item The field equations for FLRW cosmologies are second-order in time derivatives.
\end{enumerate}
Note that, as a corollary, condition (2) also ensures that general metric perturbations about maximally symmetric spaces are second-order~\cite{Hennigar:2017ego, PabloPablo3}. As such, the theories we construct will necessarily be ghost-free.

For each class of metric, the field equations can be computed from the reduced action. For the black hole ansatz
\begin{equation}
ds^2=-N^2(r)f(r) dt^2+\frac{dr^2}{f(r)}+r^2d\Omega_{(2)}\, ,
\end{equation} 
the field equations are given by
\begin{equation}
\frac{\delta S}{\delta N} = - 8 \pi r^2 {\cal E}_t{}^t   \, , \quad \frac{\delta S}{\delta f} = \frac{4 \pi r^2}{f} \left( {\cal E}_t{}^t - N {\cal E}_r{}^r \right)  \, ,
\end{equation}
while for the FLRW cosmologies,
\begin{equation}
ds^2 = -N^2(t) dt^2 + a^2(t) \left[\frac{dr^2}{1-kr^2} + r^2 d\Omega_{(2)} \right] \, ,
\end{equation}
the field equations are given by
\begin{equation}
\frac{1}{\Sigma_k}\frac{\delta S}{\delta N} \bigg|_{N=1} = -2 a^3 {\cal E}_t{}^t   \, , \quad \frac{1}{\Sigma_k}\frac{\delta S}{\delta a} \bigg|_{N=1}  = -6 a^2 {\cal E}_*{}^*  \, .  
\end{equation}
In the case of the FLRW metric, the function $N(t)$ serves as a Lagrange multiplier (to be set to unity at the end) allowing for the determination of the ${\cal E}_t{}^t$ component of the field equations from the reduced action.

At an operational level, the selection criteria are implemented in the following way. To ensure (2), we demand that $\delta S/\delta f = 0$ for a spherically symmetric black hole ansatz (and $\delta S/\delta V = 0$ for the Taub-NUT ansatz~\cite{Bueno:2018uoy}). This condition ensures there is only one independent component of the field equations, and so the black hole solutions can be characterized by a single function $f(r)$.  Condition (3) can be ensured by requiring that the Lagrangian is degenerate, \ie
\begin{equation}
\left(\frac{\partial^2 \sqrt{-g} {\cal L}}{\partial \dot{N}^2} \right) \left(\frac{\partial^2 \sqrt{-g} {\cal L}}{\partial \ddot{a}^2} \right) - \left(\frac{\partial^2 \sqrt{-g} {\cal L}}{\partial \dot{N} \partial \ddot{a}} \right)^2 = 0 \, .
\end{equation}

Of course, characterizing in complete generality the theories that meet these conditions at each order in curvature would be an enormous task due to the huge number of available invariants at high-orders in curvature~\cite{0264-9381-9-5-003}. Here we follow a simpler approach, as was done in~\cite{PabloPablo4}, constructing the higher-order densities as products of lower-order densities. We find that the following set of invariants is sufficient to generate theories at each order in curvature:
\begin{align}
&R \, , \quad Q_1 = R_{ab}R^{ab} \, , \quad Q_2 = R_{abcd} R^{abcd} \, , \quad C_1 = R_a{}^b{}_c{}^d R_b{}^e{}_d{}^f R_e{}^a{}_f{}^c \, , \quad C_2 = R_{ab}^{cd} R_{cd}^{ef} R_{ef}^{ab} \, , \quad C_3 = R_{abcd} R^{abc}{}_e R^{de} \, ,		\quad 
	\nn
&A_2 =R_{a}{}^{e}{}_{c}{}^{f} R^{abcd} R_{bjdh} R_{e}{}^{j}{}_{f}{}^{h} \, , \quad   A_{10} =  R^{ab} R_{a}{}^{c}{}_{b}{}^{d} R_{efhc} R^{efh}{}_{d} \, , \quad A_{14} =  R^{ab} R^{cd} R_{ecfd} R^{e}{}_{a}{}^{f}{}_{b} \, , \quad  
	\nn
&I_1 = R_{ce}{}^{ae} R_{af}{}^{cd} R_{gi}{}^{ef} R_{bj}{}^{gh}R_{dh}{}^{ij} \, .
\label{invars}
\end{align}
The numbering of the cubic and quartic terms is chosen to match the `standard' ordering for these invariants, (see, \eg~\cite{Aspects, Ahmed:2017jod}).

Let us now present examples of invariants that meet the conditions mentioned above. At cubic order, the result was presented in~\cite{Arciniega:2018fxj}. In terms of the basis above it can be expressed as:
\begin{equation}
{\cal R}_{(3)} = \frac{3}{16} \left[R^3 - 4 Q_1 R - Q_2 R - \frac{16}{3} C_1 - 2 C_2 + 8 C_3 \right] = -\frac{1}{8} \left({\cal P} - 8 {\cal C} \right) - \frac{1}{16} {\cal X}_6 + {\cal T}_3\, ,
\end{equation}
where ${\cal P}$ is the density corresponding to Einsteinian cubic gravity~\cite{PabloPablo}, ${\cal C}$ is a density identified in~\cite{Hennigar:2017ego}, ${\cal X}_6$ is the six-dimensional Euler density, and ${\cal T}_3$ is the term defined above Eq.~(1) in~\cite{Arciniega:2018fxj}; both of these latter terms vanish identically for four-dimensional metrics. At higher-order in curvature, the invariants are of course more complicated. We have constructed explicit examples up to tenth-order in curvature; here, for the sake of conciseness, we list examples up to eighth-order in curvature:
\begingroup
\allowdisplaybreaks
\begin{align}
{\cal R}_{(4)} =& - \frac{1}{192} \bigg[5 R^4 - 60 R^2 Q_1 + 30 R^2 Q_2 - 160 R C_1 + 32 R C_2 - 104 R C_3 + 272 Q_1^2  - 256 Q_1 Q_2 + 45 Q_2^2 -240 A_2
	\nn
	&+ 336 A_{10} + 48 A_{14}  \bigg]	
	\\
{\cal R}_{(5)} =& -\frac{1}{5760} \bigg[ 15R^{5}-36R^{3}Q_{{1}}-224R^{3}Q_{{2}}-336R^{2}C_{{1}}
-140R^{2}C_{{2}}+528R^{2}C_{{3}}-592RQ_{{1}}^{2}+1000RQ_
{{1}}Q_{{2}}
	\nn
	&+301RQ_{{2}}^{2}-912RA_{{2}}-928RA_{{10}}+1680RA
_{{14}}+1152Q_{{1}}C_{{1}}+264Q_{{1}}C_{{2}}+312Q_{{2}}C_{{2}}-
64Q_{{1}}C_{{3}}
	\nn
	&-2080Q_{{2}}C_{{3}}+4992I_{{1}}\bigg]	
	\\
{\cal R}_{(6)} =&\frac{1}{3594240} \bigg[56813R^{6}-523188R^{4}Q_{{1}}+6234R^{4}Q_{{2}}+798849R
^{3}C_{{2}}-558622R^{3}C_{{3}}+1235848R^{2}Q_{{1}}^{2}
	\nn
	&-163250R^{2}Q_{{1}}Q_{{2}}+42084R^{2}Q_{{2}}^{2}-707808R^
{2}A_{{2}}+231048R^{2}A_{{10}}+439920R^{2}A_{{14}}-5265366RQ
_{{1}}C_{{2}}
	\nn
	&+23208RQ_{{2}}C_{{2}}+4902132RQ_{{1}}C_{{3}}+44880R
Q_{{2}}C_{{3}}-704400Q_{{1}}^{3}+289200Q_{{1}}Q_{{2}}^{2}-
62400Q_{{2}}^{3}+1168128RI_{{1}}
	\nn
	&+792000Q_{{1}}A_{{2}}+374400
Q_{{2}}A_{{2}}-723600Q_{{2}}A_{{10}}-676800{C_{{1}}}^{2}+7903368
C_{{1}}C_{{2}}-8581680C_{{1}}C_{{3}}-3782484C_{{2}}^{2}
	\nn
	&+15454692 C_{{2}}C_{{3}}-12753720C_{{3}}^{2}
 \bigg]	
	\\
{\cal R}_{(7)} =& -\frac{1}{1692903628800} \bigg[ 440336334R^{5}Q_{{1}}+225293473325R^{4}C_{{2}}-156064142950{
R}^{4}C_{{3}}-77884756172R^{3}Q_{{1}}Q_{{2}}	
	\nn
	&+190397473375R^{3}Q_{{2}}^{2}+399100531608R^{2}Q_{{2}}C_{{1}}-1367252157606R^{
2}Q_{{1}}C_{{2}}-102580664328R^{2}Q_{{2}}C_{{2}}
	\nn
	&+1109980123860R^{2}Q_{{1}}C_{{3}}-675582698988R^{2}Q_{{2}}C_{{3}}-637404411108
RQ_{{1}}Q_{{2}}^{2}+83691964836RQ_{{2}}^{3}
	\nn
	&+1044160568064R^{2}I_{{1}}+239509768896RQ_{{1}}A_{{2}}-228185841312RQ_{{2}}A_{{2}}+
668225357808RQ_{{1}}A_{{10}}
	\nn
	&+476215786872RQ_{{2}}A_{{10}}+
976705691136RC_{{1}}^{2}+1470463896712RC_{{1}}C_{{2}}-
1279315669424RC_{{1}}C_{{3}}
	\nn
	&-976616255780RC_{{2}}^{2}+
3692325167284RC_{{2}}C_{{3}}-1168144584648RC_{{3}}^{2}+
59581564112Q_{{1}}Q_{{2}}C_{{1}}
	\nn
	&+515811286184Q_{{2}}^{2}C_{{1}}-
225367377256Q_{{1}}Q_{{2}}C_{{2}}-267119623780Q_{{2}}^{2}C_{{2}}
+1862849653896Q_{{1}}Q_{{2}}C_{{3}}
	\nn
	&+492136418316Q_{{2}}^{2}C_{{3
}}+89993670912C_{{1}}A_{{2}}-11491451136C_{{2}}A_{{2}}+
839177549568C_{{3}}A_{{2}}
	\nn
	&-1892820532608C_{{1}}A_{{10}}+
2478458228544C_{{2}}A_{{10}}-7514929844352C_{{3}}A_{{10}}-
3604473976320C_{{1}}A_{{14}}
	\nn
	&-4636547758080Q_{{1}}I_{{1}}-
80412922368Q_{{2}}I_{{1}}
\bigg]	 
	\\
{\cal R}_{(8)} =& -\frac{1}{55678707292029419520}\bigg[
457128401388753177R^{8}-5733790494231545862R^{6}Q_{{1}}
	\nn
	&-1545850314903381546R^{6}Q_{{2}}-206671431127697838R^{5}C_{{2}}
+99354249720411072R^{5}C_{{3}}
	\nn
	&+78534617625715894344R^{4}Q_{{1}}^{2}-12692968076910280123R^{4}Q_{{2}}^{2}-2399008460254826784
R^{4}A_{{2}}
	\nn
	&+95396027232989898768R^{4}A_{{10}}-
219511908936538092576R^{4}A_{{14}}-308888884222030865784R^{2}{
Q_{{1}}}^{3}
	\nn
	&+42487096725733757568R^{2}Q_{{1}}^{2}Q_{{2}}+
56638770909883900164R^{2}Q_{{1}}Q_{{2}}^{2}+760431228416527536
R^{2}Q_{{2}}^{3}
	\nn
	&+272125034870687008128R^{3}I_{{1}}-
2004936678226190880R^{2}A_{{2}}Q_{{2}}-495008060669893397760R^
{2}A_{{10}}Q_{{1}}
	\nn
	&+1106432545771753207776R^{2}A_{{14}}Q_{{1}}-
233878181709100032R^{2}C_{{2}}^{2}+3211277274960475680R^{2}C
_{{2}}C_{{3}}
	\nn
	&-58580698119180073872R^{2}C_{{3}}^{2}-
81583684274155268720RC_{{1}}Q_{{2}}^{2}-7154443498022983512RC_{{
2}}Q_{{1}}^{2}
	\nn
	&+40946886168028287256RC_{{2}}Q_{{2}}^{2}+
349032350277790847136RC_{{3}}Q_{{1}}^{2}-125918130177013828008RC
_{{3}}Q_{{2}}^{2}
	\nn
	&+419234950034409600Q_{{1}}^{4}-
50518172239870608Q_{{1}}Q_{{2}}^{3}+9911637028991880Q_{{2}}^{4
}+43644245348840428032RA_{{2}}C_{{1}}
	\nn
	&-22360397876623103232RA_{{2}}
C_{{2}}+75630694229824093056RA_{{2}}C_{{3}}-14841840827791902240RA
_{{10}}C_{{2}}
	\nn
	&+685845113753249942976RA_{{10}}C_{{3}}+
41631119399244856032RA_{{14}}C_{{2}}-1405506700276515250560RA_{{14
}}C_{{3}}
	\nn
	&-1620171054437554826496RI_{{1}}Q_{{1}}-769587637329838080
RI_{{1}}Q_{{2}}-622703543044197888A_{{2}}Q_{{1}}Q_{{2}}
	\nn
	&+
10633927339526400A_{{2}}Q_{{2}}^{2}+216123540558149952A_{{10}}{Q
_{{2}}}^{2}+2373774922358880768C_{{1}}^{2}Q_{{1}}
	\nn
	&+175772732137475289600C_{{1}}^{2}Q_{{2}}-168361120304786722656C_{
{1}}C_{{2}}Q_{{2}}+297036572725084407360C_{{1}}C_{{3}}Q_{{2}}
	\nn
	&+40305033606596700144C_{{2}}^{2}Q_{{2}}-138497624651063373744C_{{
2}}C_{{3}}Q_{{2}}+53160989298451803360C_{{3}}^{2}Q_{{2}}
	\nn
	&-407876360805688320A_{{2}}^{2}+389927731483809792A_{{2}}A_{{10}}+
2119740796344745663488C_{{1}}I_{{1}}
	\nn
	&-1058014677342964981248C_{{2}}
I_{{1}}+3173100906526796818944C_{{3}}I_{{1}}
 \bigg] 	 \, .
\end{align}
\endgroup

As emphasized in the main text, within the set of possible Lagrangians generated by the set of invariants~\eqref{invars}, and certainly more generally, there exist multiplicities. That is, there exist distinct densities ${\cal R}_{(n)}^A$ and ${\cal R}_{(n)}^B$ that differ by a density ${\cal T}_{(n)}^{AB}$ with the property that ${\cal T}_{(n)}^{AB}$ makes no contribution to the field equations for the three classes of metrics considered here. Interestingly, we find that this appears to be the \textit{only} type of degeneracy possible. That is, we find no examples of densities ${\cal R}_{(n)}^A$ and ${\cal R}_{(n)}^B$ satisfying our conditions that make distinct contributions to the field equations for  the classes of metrics considered here. 

An important feature for future study is to distinguish the densities ${\cal T}_{(n)}^{AB}$ that are genuinely trivial for four-dimensional geometries from those that are trivial only on the restricted classes of metrics considered here. Invariants that vanish only for our restricted selection of metrics could contribute in less symmetric scenarios, \eg in the study of metric perturbations. Further consistency checks would be required to select the ``correct'' combinations from this larger set. A preliminary analysis of (scalar) cosmological perturbations including terms up to sixth-order in curvature suggests that, so long as the conditions advocated above are met, the equations of motion for the perturbations are always second-order in time derivatives. However, we find that the precise form of the perturbation equations \textit{is} sensitive to the particular representative ${\cal R}_{(n)}^A$ chosen, provided $n \ge 5$.

\section{Schwarzschild-like black holes}
\label{sec:schwHoles}

As mentioned in appendix~\ref{densis}, the theory possesses non-hairy black hole solutions of the form 
\begin{equation}
ds^2=-f(r)dt^2+\frac{dr^2}{f(r)}+r^2d\Omega_{(2)}\, ,
\end{equation}
\noindent
where $f$ satisfies \cite{PabloPablo4}
\begin{align}\label{eqf}
&-(f-1)r-\sum_{n=3}^{\infty}\frac{n(n-1)\lambda_n}{2^n}L^{2n-2}\left(\frac{f'}{r}\right)^{n-3}\Bigg[\frac{f'^3}{n}+\frac{(n-3)f+2}{(n-1)r}f'^2-\frac{2}{r^2}f(f-1)f' \qquad\qquad\qquad \\ \notag
&\qquad\qquad\qquad\qquad\qquad\qquad\qquad\qquad\qquad\qquad\qquad -\frac{1}{r}f f''\left(f'r-2(f-1)\right)\Bigg] =2GM+\frac{1}{3}\Lambda r^3\, .
\label{kfequationn2}
\end{align}
%
The solutions to this equation have a number of remarkable properties that we shall briefly review in this appendix. 

The field equation does not admit an exact solution, but the important features can be understood by considering near horizon and asymptotic solutions. To leading order, the asymptotic solution reads
\begin{equation}
f(r) = 1 - \frac{2 G M}{r} - \frac{27 \lambda_3 L^4 (GM)^2}{r^6} + f_{\rm h}(r) + {\cal O}(r^{-7})\, ,
\end{equation}
where $f_{\rm h}(r)$ is the homogeneous solution:
\begin{equation}
f_{\rm h}(r) \approx A r^{1/4} \exp\left[\frac{4 r^{5/2}}{15 L^2 \sqrt{2   M\lambda_3}} \right] + B r^{1/4} \exp\left[\frac{-4 r^{5/2}}{15 L^2 \sqrt{2 G M\lambda_3}} \right] \, .
\end{equation}
From this we deduce that $A = 0$  and $\lambda_3 > 0$ are requirements for the existence of positive mass asymptotically flat solutions. The first of these conditions fixes one of the two integration constants characterizing the solution. If the second of these conditions were violated, the solution would oscillate, rather than decay, near infinity. Here we have only considered the leading term in the asymptotic expansion, assuming $\lambda_3 \neq  0$. The general result is that the first non-vanishing coupling $\lambda_n$ must be positive.

A near-horizon solution can be obtained by substituting 
\begin{equation}
f_{\rm nh} (r) = 4 \pi T (r- \rh) + \sum_{n=2}^\infty a_n (r-\rh)^n
\end{equation}
into the field equations and solving order-by-order for the unknown coefficients. The first two of these relations read:
\begin{align}
2 G M &= \rh - \sum_{n=3}^\infty \frac{\lambda_n L^{2(n-1)} (4\pi T)^{n-1}}{2^n \rh^{n-2}} \left(2n + (n-1) 4\pi T\rh \right)  \, , \label{nh1}
	\\
0 &= - 1 + 4 \pi T \rh +   \sum_{n=3}^\infty \frac{\lambda_n L^{2(n-1)} (4\pi T)^{n-1}}{2^n \rh^{n-1}} \left(2n + (n-3) 4\pi T\rh \right) 	\, . \label{nh2}
\end{align}
At the next order in $(r-\rh)$ the parameters $a_2$ and $a_3$ appear, with the next parameter then showing up (linearly) at each higher-order. The parameters $a_n$ for $n \ge 3$ can be solved for in terms of $a_2$ and the couplings. Thus the entire near horizon solution is specified by two parameters: the mass and  $a_2$. The value of $a_2$ itself can be determined using either numerical methods or analytical approximations to join the near horizon solution on to the asymptotic expansion. These methods reveal that there is a unique value of $a_2$ that ensures the two regions connect smoothly together --- see, for example,~\cite{PabloPablo4, Hennigar:2018hza} for more explicit details. It is in this sense that  the black holes \textit{have no hair}: the two integration constants characterizing~\eqref{eqf} have been fixed by demanding the solution is asymptotically flat, the only free parameter is the mass.

The first two near-horizon equations~\eqref{nh1} and \eqref{nh2} determine $M$ and $T$ in terms of $\rh$ and the couplings, indicating that the \textit{thermodynamic properties of the black holes can be studied fully analytically}. The thermodynamic properties of large black holes are very similar to the usual Schwarzschild solution, but the behavior of small black holes is markedly different.  When the higher-curvature terms are non-trivial, there exists a mass scale $M_{\rm stable}$ where the heat capacity changes sign and the \textit{black holes are thermally stable}.  As discussed in~\cite{PabloPablo4}, the black holes with masses smaller than $M_{\rm stable}$ will never completely evaporate. It is a tantalizing possibility that these small stable black holes could be responsible for dark matter. Further, preliminary phenomenological studies involving black hole shadows and gravitational lensing have suggested mechanisms by which the couplings of the theory may be constrained~\cite{Hennigar:2018hza, Poshteh:2018wqy}.


In practice, the scale  $M_{\rm stable}$ can be determined by finding a simultaneous solution to~\eqref{nh2} and
\begin{equation}
0 = -1 + 2 \pi T \rh +  \sum_{n=3}^\infty  \lambda_n \left[(1 - 2 \pi T \rh)  + \frac{n(n-1) }{12 \pi T \rh} \left(1 + 2 \pi T \rh  \right)^2\right] \left( \frac{2 \pi T L^2}{\rh } \right)^{n-1}  \, ,
\end{equation}
(which is just the condition determining a maximum for the temperature) and then plugging the resulting values for $T$ and $\rh$ back into~\eqref{nh1} to determine the mass, $M_{\rm stable}$. For the models considered in the main text, the determination of $M_{\rm stable}$ reduces to solving a system of transcendental equations. 



Lastly, let us note that many of the remarkable properties of four-dimensional asymptotically flat black holes hold more generally. For example, the same theory that yields the black hole solutions described in this appendix also admits Taub-NUT/bolt solutions~\cite{Bueno:2018uoy}. Those metrics are once again non-hairy generalizations of the Einstein gravity Taub-NUT/bolt solutions and the thermodynamics can be studied exactly. Similar features apply for higher-dimensional generalizations of the theory~\cite{Hennigar:2017ego, PabloPablo3, Ahmed:2017jod}.



\bibliographystyle{apsrev4-1} 
\vspace{1cm}
\bibliography{Gravities} 

\end{document}